\documentclass[10pt,journal,final,twocolumn]{IEEEtran}
\usepackage{algorithm}
\usepackage{algpseudocode}
\usepackage{amsfonts}
\usepackage{bm}
\usepackage{amsmath}
\usepackage{cite}
\usepackage{float}
\usepackage{stfloats}

\usepackage{amssymb}
\usepackage{enumerate}
\usepackage{color,xcolor}
\usepackage{colortbl}
\usepackage{graphicx}
\usepackage{multirow,tabularx}
\usepackage{subfigure}
\usepackage{hyperref}

\makeatletter

\newcommand{\Rmnum}[1]{\expandafter\@slowromancap\romannumeral #1@}
\makeatother
\usepackage{bookmark}
\usepackage{soul}
\setstcolor{blue}

\ifCLASSINFOpdf

\else

\fi

\newcolumntype{I}{!{\vrule width 1.0pt}}
\newlength\savedwidth
\newcommand\whline{\noalign{\global\savedwidth\arrayrulewidth
                            \global\arrayrulewidth 1.0pt}%
                   \hline
                   \noalign{\global\arrayrulewidth\savedwidth}}
\newlength\savewidth

\definecolor{lightgray}{RGB}{204,204,204}
\definecolor{lightblue}{RGB}{204,255,205}
\definecolor{lightgreen}{RGB}{255,255, 204 }
\definecolor{columbiablue}{rgb}{0.61, 0.87, 1.0}
\definecolor{hblue}{RGB}{255,204,153}
\definecolor{aquamarine}{RGB}{0.5, 1.0, 0.83}
\usepackage{multirow}

\begin{document}

\title{\huge{Distributed Task-Oriented Communication Networks with Multimodal Semantic Relay and Edge Intelligence}}

\author{
Jie Guo,~\IEEEmembership{Member,~IEEE,}
       Hao Chen,\IEEEmembership{}
       Bin~Song,~\IEEEmembership{Senior Member,~IEEE,}\\
       Yuhao Chi,~\IEEEmembership{Member,~IEEE,}
       Chau Yuen,~\IEEEmembership{Fellow,~IEEE,}
	   Fei Richard Yu,~\IEEEmembership{Fellow,~IEEE,} \\
	     Geoffrey Ye Li,~\IEEEmembership{Fellow,~IEEE},
      and Dusit Niyato,~\IEEEmembership{Fellow,~IEEE}
\thanks{Jie Guo, Bin Song, and Yuhao Chi are with the State Key Laboratory of Integrated Services Networks, Xidian University, Xi'an, 710071, China.}
\thanks{Hao Chen is with the Hangzhou Institute of Technology, Xidian University, Hangzhou, 311231, China.}
\thanks{Chau Yuen and Dusit Niyato are with the School of Electrical and Electronics Engineering and the School of Computer Science and Engineering respectively, Nanyang Technological University, 639798, Singapore.}
\thanks{Fei Richard Yu is with the Department of Systems and Computer Engineering, Carleton University, Ottawa, K1S 5B6, Canada.}
\thanks{Geoffrey Ye Li is with the Faculty of Engineering, Imperial College London, London, SW7 2AZ, UK.}
}

\maketitle

\begin{abstract}
In this article, we present a novel framework, named distributed task-oriented communication networks (DTCN), based on recent advances in multimodal semantic transmission and edge intelligence. In DTCN, the multimodal knowledge of semantic relays and the adaptive adjustment capability of edge intelligence can be integrated to improve task performance. Specifically, we propose the key techniques in the framework, such as semantic alignment and complement, a semantic relay scheme for deep joint source-channel relay coding, and collaborative device-server optimization and inference. Furthermore, a multimodal classification task is used as an example to demonstrate the benefits of the proposed DTCN over existing methods. Numerical results validate that DTCN can significantly improve the accuracy of classification tasks, even in harsh communication scenarios (e.g., low signal-to-noise regime), thanks to multimodal semantic relay and edge intelligence.


\end{abstract}

\IEEEpeerreviewmaketitle
\section{Introduction}

With the advances in artificial intelligence (AI), the Internet of Things (IoT), and communication technologies, the transmitted data types and supported multimedia applications have grown increasingly diverse. Edge devices (e.g., base stations, sensors) are becoming widely deployed, and sensory data is easily accessed. However, with the explosive growth of bandwidth-consuming services in the future, such as extended reality (XR), intelligent vehicular networks, and smart cities, wireless bandwidth will be insufficient and transmission latency will increase. To address these issues, researchers are looking for breakthroughs in communication technologies, one of which is task-oriented communication~\cite{Semanticoverview}.

Compared with conventional data-oriented communication, which attempts to recover every single bit accurately, task-oriented communication, which only transmits task-related semantic information, is more effective. For example, for gender classification in images, only gender-related features are useful, while other detailed features, such as hair color and facial expressions, are superfluous. Moving targets in video surveillance tasks often convey more task-relevant information and should receive more attention than the background. Instead of focusing on accurate data recovery, task-oriented communication extracts only minimal but sufficient information for the inference task, which has the potential to greatly improve the efficiency of communication.

At the front end of task-oriented communication networks, the data formats of different sources are diverse, including texts, images, videos, etc., but they contain tremendously correlated contextual information. To reduce redundancy and latency in transmission, correlations among multimodal data sources can be exploited. In \cite{multi_gat}, a multimodal mixture-of-experts model is designed to disentangle and extract salient modality-specific features that enable feature interactions, and a message-passing-based graphical attention approach is introduced to capture multimodal semantic correlation. Ding \emph{et al.} ~\cite{MuKEA} have explored an explicit triplet to represent multimodal knowledge, which correlates visual objects and factual answers with implicit relations. In the field of multimodal learning, multimodal feature extraction and aggregation techniques have been extensively investigated. However, traditional multimodal learning techniques cannot be directly applied to task-oriented communication networks without accounting for the impact of wireless communication factors (e.g., wireless channels, communication resources, and transmission latencies)\cite{Learning}. Besides, existing task-oriented communication networks do not fully consider the latent correlation between multimodal information, resulting in the underutilization of the source data and low task performance.


To achieve reliable transmission of multimodal data,  conventional data-oriented communication often employs source-channel separation coding, which is commonly designed with the assumption of infinite code length. However, in distributed task-oriented semantic communication scenarios, deep joint source-channel coding (JSCC) is actually superior to separating source and channel coding for semantic features that are usually  finite-length, and their transmissions often demand stringent latency. Recently,  deep JSCCs, such as DeepSC~\cite{deepscarticle}, transmit semantic features extracted by neural networks, and the semantic receiver performs relevant tasks using semantic features instead of reconstructing source information. Nevertheless, most semantic communications nowadays only consider the end-to-end (E2E) communication paradigm and do not exploit wireless relays. Actually, edge servers can act as relays for data amplification and semantic complement, thus improving communication stability against noise.

\begin{figure*}[htbp]\vspace{-0.3cm}
	\centering
	\includegraphics[width=0.85\textwidth]{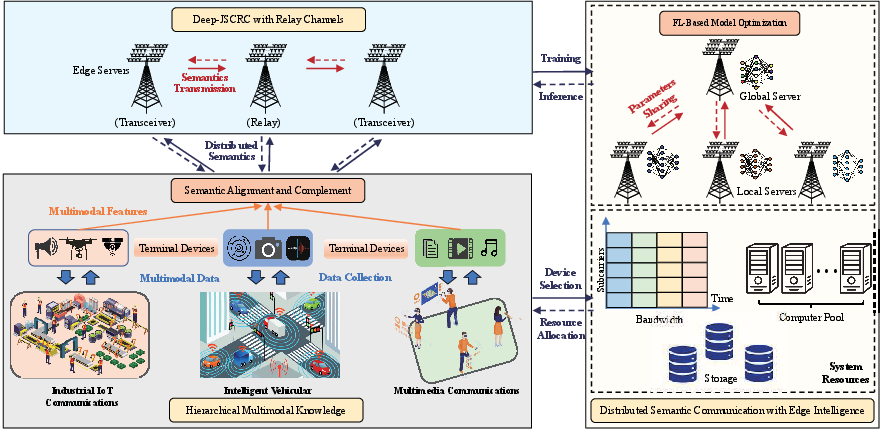}\vspace{-0.2cm}
	\caption{Illustration of the proposed DTCN framework.}
	\label{dtcn}
\end{figure*}

Due to the real-time requirement of specific data services, it is obviously unrealistic to transmit all data to a cloud computing center. Edge intelligence, which integrates AI and edge computing, provides a promising solution \cite{Shannon}. Deep neural networks (DNNs) are deployed on edge servers that are distributed in different geographic locations. The potential correlation of tasks performed by edge servers in different locations can be exploited by semantic communication for significant performance improvement. Mills \emph{et al.} \cite{fl_dis} have investigated multi-task federated learning (FL) in wireless communications, where each edge device trains two local models of different tasks and the base station (BS) aggregates the global model. Therefore, compared to conventional distributed communication, tasks can be completed with low latency by leveraging the correlation of multimodal information to supplement and strengthen the semantic information.



Based on the above analyses, both multimodal semantic relay and edge intelligence should be considered for highly efficient task-oriented communications. To this end, we propose a novel distributed semantic learning-based framework, termed distributed task-oriented communication networks (DTCN). The key of DTCN is to distribute and fuse hierarchical multimodal knowledge with the aid of mutlimodal semantic relay and leverage edge intelligence for various tasks. For the DTCN framework, we introduce key techniques for hierarchical multimodal knowledge, a semantic relay scheme for deep joint source-channel relay coding (Deep-JSCRC) and collaborative device-server optimization and inference. The proposed DTCN outperforms existing methods for multimodal classification experiment using the UPMC Food-101 dataset~\cite{dataset}. Even with missing modality and data corruption, DTCN can still complete the task accurately.

\section{DTCN Framework}
\label{sec:intel}
This section introduces the proposed DTCN framework. As shown in Fig.~\ref{dtcn}, it incorporates hierarchical multimodal knowledge, {Deep-JSCRC} with relay channels, and distributed semantic communication with edge intelligence. Specifically, hierarchical multimodal knowledge is developed to acquire and align multimodal features as well as capture the correlation between multimodal information. The {Deep-JSCRC} with relay channels complements the necessary multimodal features to perform the tasks and transfers multimodal knowledge from the transmitter to the receiver. The distributed semantic communication network accomplishes specific tasks collaboratively by utilizing heterogeneous computing capabilities. The contributions, key technologies, and desired properties of the DTCN framework are summarized in Table~\ref{inno}.

\newcommand{\tabincell}[2]
{\begin{tabular}{@{}#1@{}}#2\end{tabular}}
\begin{table*}[h]\scriptsize\vspace{-0.5cm}
	\renewcommand{\arraystretch}{1.3}
	\caption{Contributions, key technologies, and desired properties of DTCN framework.
	}\label{com_obj}\vspace{-0.2cm}
	\centering
    \scalebox{0.85}{
	\begin{tabular}{IcI>{\columncolor{lightgreen}}lI>{\columncolor{lightblue}}lI >{\columncolor{columbiablue}}lI}
		\whline
		\scriptsize \cellcolor{lightgray}{\textbf{Approaches}} & \scriptsize \cellcolor{lightgray}{ \centering{\textbf{Contributions compared with conventional methods \cite{MuKEA,Learning,deepscarticle,Shannon}}} }& \scriptsize \cellcolor{lightgray}{\centering{\textbf{Key technologies}}}& \scriptsize \cellcolor{lightgray}{\centering{\textbf{Desired properties}}}  \\
		\whline
		{\cellcolor{hblue} \tabincell{c}{\vspace{1.3cm}\textbf{Hierarchical  }\vspace{-1.2cm}\\ \textbf{multimodal knowledge}}} & \tabincell{l}{ $\bullet$ Extract multiscale features from multimodal data \vspace{0.5mm}\\$\bullet$ Aggregate interactions of multiscale semantic features  \vspace{0.1cm} \\$\bullet$ Explore the semantic correlations of multimodal data} &\tabincell{l}{$\bullet$ Multi-GAT: Establish semantic correlation \\\quad between multiscale multimodal information} & \tabincell{l}{ $\bullet$ Comprehensive semantic features \vspace{0.5mm}\\ 
		$\bullet$	Hierarchical semantic alignment \vspace{0.5mm}\\
		$\bullet$ Performance improvement of multimodal tasks}
        \\
		\whline
		{\cellcolor{hblue} \tabincell{c}{\vspace{1.25cm} \textbf{Deep-JSCRC}\vspace{-1.2cm}\\ \textbf{with relay channels}}} & \tabincell{l}{$\bullet$ Semantic-aware wireless relays \\ $\bullet$ Multimodal background knowledge}& \tabincell{l}{$\bullet$ {Deep-JSCRC: Multimodal information fusion  }\\\quad {based on multimodal semantic correlation}}& \tabincell{l}{$\bullet$ Wireless and semantic noise suppression \vspace{0.5mm}\\ $\bullet$ Semantic complementation and strengthening}\\
		\whline
		{\cellcolor{hblue}\tabincell{c}{\vspace{1.3cm}\textbf{Distributed semantic} \vspace{-1.2cm}\\\vspace{0.4cm}\textbf{communication with
				}\vspace{-0.3cm}\\ \textbf{ edge intelligence}}} & \tabincell{l}{ $\bullet$ 
                {Resource allocation based on semantic contributions}\\$\bullet$ Adaptive workload allocation for distributed devices \vspace{0.75mm}\\$\bullet$ Collaborative learning and inference with privacy protection}&
        \tabincell{l}{
			$\bullet$ Collaborative device-server optimization \\ ~ and inference: semantic resource allocation,\\ ~  distributed learning, and joint inference
            }&
		\tabincell{l}{
            $\bullet$ Efficient resource usage and lower latency
            \vspace{0.5mm}\\
			$\bullet$ Adaptive workload balancing \vspace{0.5mm}\\ 
			$\bullet$ Training acceleration with privacy protection
           } \\
		\whline	
	\end{tabular}
 }
\label{inno}
\end{table*}

\vspace{-0.3cm}
\subsection{Hierarchical Multimodal Knowledge}
Hierarchical multimodal knowledge consists of multiscale multimodal feature extraction and heterogeneous multimodal graphs, which are employed to establish task-specific full-scale perception and semantic information correlation of multimodal data, respectively. The details are described as follows.
\subsubsection{Multiscale Multimodal Feature Extraction}
In general, it is inadequate to accomplish most tasks by employing only fixed-scale features of multimodal data. For example, for image-text retrieval tasks, most of the existing works only focus on local relationship alignment but ignore the importance of global features as contextual information for accurate semantic matching. In contrast, multiscale feature extraction methods for multimodal data aim to establish comprehensive perceptions of specific tasks. Therefore, to enable efficient task-oriented communications, hierarchical semantic alignment of multimodal data for communication tasks is performed by multilevel neural network layers, while graph attention networks are employed to aggregate feature interactions of multiscale semantic information.




\subsubsection{Heterogeneous Multimodal Graph}
A heterogeneous multimodal graph is created by integrating complementary multimodal information from various perspectives to fully characterize the properties of objects acquired by distributed devices.
The multimodal features can be encoded by three layers of graphs, which include a visual scene graph, a semantic graph, and a knowledge graph (KG). For images and videos, the scene graph is constructed from contextual information about the objects, including visual objects, the spatial and subordinate relationships between them, etc., unlike conventional object detection methods. Meanwhile, high-level information obtained from text data and other modalities provides additional semantic information. The semantic graph is developed for semantic inference by establishing the association between multiple objects and semantic information in multimodal data. KG, which consists of a large number of nodes and edges, provides side information and semantic correlation for improving task performance. 
\vspace{-0.2cm}
\subsection{Deep-JSCRC with Relay Channels}
In task-oriented communications, due to the heterogeneity of user attributes and the diversity of tasks, multimodal background knowledge among different tasks is diverse, which leads to the inconsistency of multimodal semantic information between the transmitter and receiver, i.e., semantic noise. As a result, transmission errors caused by semantic noise are a common and challenging problem in communications that are difficult to suppress using conventional communication techniques. Therefore, exploiting the inherent correlation between multimodal information to strengthen semantics and overcome semantic noise is a promising area of research. However, if multimodal information correlation-based semantic complement and target tasks are performed concurrently at the receiver, it will introduce new challenges: 1) Extremely difficult semantic complementation, especially when information may be completely lost during transmission over long distances or in harsh scenarios (e.g., low signal-to-noise regimes), making multimodal semantic correlation difficult to exploit effectively; 2) Excessive computational and storage resources as well as complex multi-task scheduling {must be provided by the receiver, which is particularly burdensome and frequently causes high transient loads.}

To address the above issues, we develop the Deep-JSCRC with the aid of semantic relays. Specifically, edge servers act as relays, providing additional knowledge and complementary semantic features to help the receiver complete the task while mitigating the interference of semantic noise. Meanwhile, multiple deployed relays can collaborate to handle the receiver's multitasking multimodal workload by sharing background knowledge and supplementing the transmitted semantics. This allows the receiver to focus on task completion and also assist distributed devices in transmitting only small necessary semantic information, lowering the cost of large-scale devices. In a nutshell, the Deep-JSCRC not only exploits the semantic correlation between multimodality to overcome semantic noise but also simplifies the design of distributed task-oriented communication systems to support multimodal task implementation at the receiver more efficiently.

\begin{figure*}[htbp]\vspace{-0.4cm}
	\centering
	\includegraphics[width=0.7\textwidth]{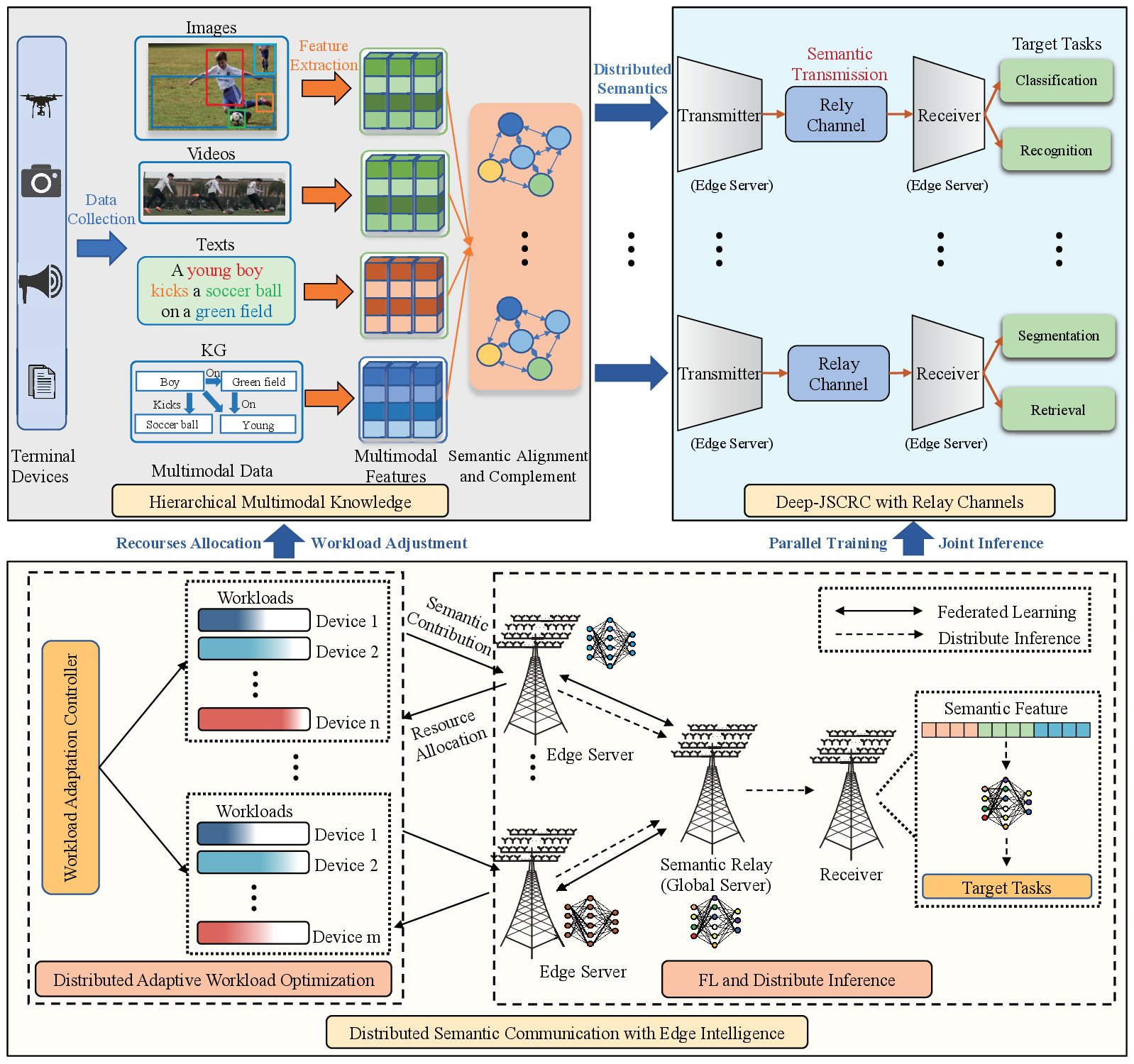}\vspace{-0.2cm}
	\caption{Techniques used in the proposed DTCN framework.}
	\label{techin}
\end{figure*}
\vspace{-0.4cm}
\subsection{Distributed Semantic Communication with Edge Intelligence}
With the rapid development of edge devices and the demand for massive data transmission, distributed semantic communication with edge intelligence is essential to overcome the limitations of communication resources and latency that are difficult to satisfy by conventional centralized communication systems. Thus, the application scenarios of distributed semantic communication are very wide, such as industrial IoT networks, intelligent vehicle networks, and multimedia communications. At the same time, it also brings new challenges, such as severe communication latency and excessive local workload. Therefore, new technologies need to be developed to overcome these issues.

\subsubsection{Resource Allocation and Device Selection}
In DTCN, multiple devices often require extracting locally transmitted semantic feature representations. For massive devices with limited resources, communication latency (e.g., queuing time) should be reduced to realize real-time training and updating of the network model. For example, edge servers will prioritize devices with high contributions to send updated model parameters and prohibit devices with low update frequencies. Furthermore, a resource allocation and device selection strategy can be employed for devices within a cluster based on workload and task contribution.

\subsubsection{Federated-Learning-Based Model Optimization} 
Note that network model updates for large-scale semantic communication systems will exhaust a large number of communication resources. As a result, the key to leveraging edge intelligence is to design mechanisms for model parameter transfer and sharing between local devices and edge servers. Recently, FL has been employed to share model parameters and transfer updated model parameters across multiple devices instead of all raw data\cite{FLarticle}, which greatly saves wireless resources, improves model convergence, and protects user privacy to the maximum extent. 

In the FL of the DTCN framework, the Deep-JSCRC is viewed as a local or global shared model. Multiple edge servers update the parameters of the local model and transmit them to the semantic relay, which is regarded as a global server. Based on these collected local models, a global model is acquired and then broadcast to multiple edge servers. The edge and global servers alternately update their own models until certain convergence criteria are satisfied. With the aid of FL, the proposed DTCN can be computed in parallel to improve training efficiency and protect user privacy.

\section{Techniques for DTCN Framework}
\label{sec:tech}
The key of DTCN is to employ hierarchical multimodal knowledge to complete certain tasks, even in cases of missing partial data. To realize effective DTCN, several fundamental techniques are  necessary. As shown in~Fig.~\ref{techin}, we classify these techniques into three categories: semantic alignment and complement, semantic relay scheme for Deep-JSCRC, and collaborative device-server optimization and inference.

\vspace{-0.2cm}
\subsection{Semantic Alignment and Complement}
Due to the difference in data representation, it is often complicated to infer the latent global (e.g., images and sentences) and local (e.g., objects in images and words in sentences) semantic correlations among different modalities. Furthermore, once fully explored, the correlated semantic information will constitute the elaborately aligned features to complement missing modalities, which is extremely important in practical communication scenarios. For this purpose, two types of techniques, including multimodal feature extraction and multimodal knowledge aggregation, should be considered.

\subsubsection{Multimodal Feature Extraction}
To extract comprehensive intramodal features from heterogeneous sensory data (e.g., images, videos, texts, multimodal KG, etc.), different algorithms should be employed. Generally, for image modality, the semantic extractor can adopt Vision Transformer (ViT) dividing the input image into several patches and sending them to Transformer blocks to generate feature representations. As for text modality, by concurrently conditioning on the bidirectional context in all layers, Bidirectional Encoder Representations from Transformer (BERT) can be used to pretrain deep representations from unlabeled text and extract context features. In addition, a multimodal knowledge encoder can be applied to extract the head entity, relationship, and tail entity representations, respectively, to form a knowledge triplet from the knowledge base. By introducing and fine-tuning strong feature extraction networks according to the task requirements, DTCN can extract more accurate and useful semantic information from heterogeneous data.

\subsubsection{Multimodal Knowledge Aggregation}
To obtain more underlying spatiotemporal features, heterogeneous multilayer graphs, which include visual graphs, semantic graphs, and KGs, are considered. A multilayer graph that consists of different types of nodes and edges can discover and aggregate the spatiotemporal correlation of different modalities. The multilayer graph can be denoted as $G=(V, E)$, where the node set $V$ represents entities and $E$ is the edge set (relationship) among object nodes passing entity-to-entity messages.
Then, a heterogeneous graph neural network based on the multimodal Graph Attention Network (Multi-GAT)\cite{multi_gat} is used to capture intermodal latent relationships for aggregating complementary multimodal knowledge. To be specific, different modality information is aligned by Multi-GAT based on semantic similarity instead of simply concatenating all multimodal information. Each modality interacts with other modalities using the attention mechanism to determine edge-weight relationships and generate intermodal complementary features. As a result, fused multimodal knowledge can be obtained by aggregating intramodal and intermodal features.


\begin{figure*}[t]\vspace{-0.3cm}
	\centering
	\includegraphics[width=0.7\textwidth]{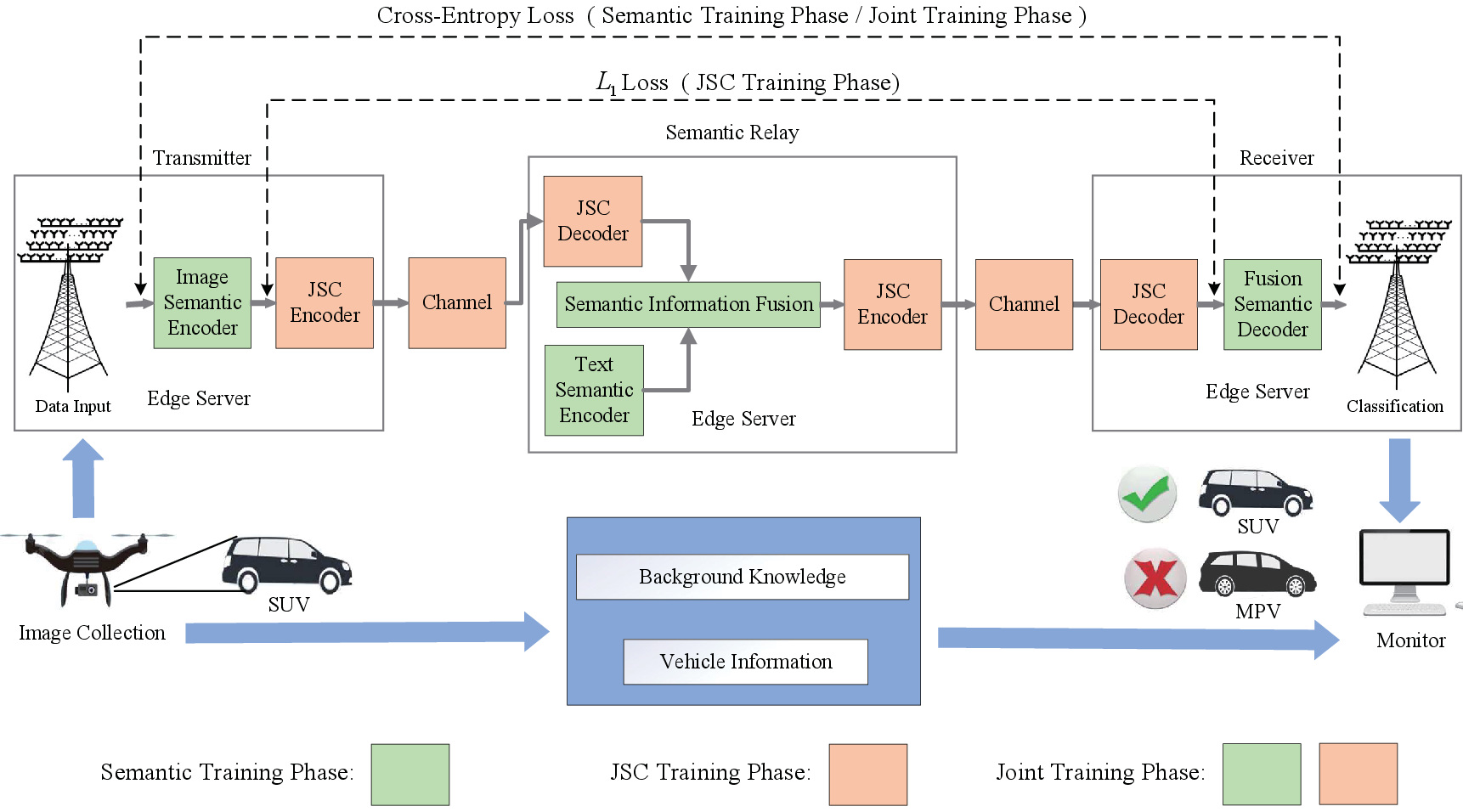}\vspace{-0.2cm}
	\caption{An example of vehicle recognition using the semantic relay scheme for Deep-JSCRC.}
	\label{relay}
\end{figure*}

\subsection{Semantic Relay Scheme for Deep-JSCRC}
In practical applications,  sharing semantic background knowledge between transmitters and receivers presents significant challenges. It is also difficult to transmit information reliably in a harsh environment. As a result, the Deep-JSCRC with rich background knowledge leverages multimodal information correlation to supplement and strengthen semantic information, mitigating the impact of wireless channel and semantic noise to improve task performance.

The Deep-JSCRC consists of a transmitter, a semantic relay, and a receiver. At the transmitter, semantic features of the image are extracted by a semantic encoder and mapped into transmission symbols delivered to the channels by a joint source channel (JSC) encoder. When the semantic feature is received and recovered by a JSC decoder at the relay, it is fused with the additional background knowledge semantics to generate enhanced semantic feature. Following a JSC encoder, the fused semantic feature is transmitted from the relay to the receiver and then recovered to complete the target task based on the user-specific requirements.

In the meantime, the Deep-JSCRC can significantly protect user privacy over traditional decode-and-forward relay schemes. The main reason is that the Deep-JSCRC concentrates on semantic-level complementation and enhancement without reconstructing the user's original data, and the process of acquiring task-oriented semantic features based on DNN is not always fully reversible. As a result, even if attackers can successfully steal the task-oriented semantic features transmitted by the semantic relay, accessing the explicit tasks of the receiver and specific background knowledge is extremely difficult. It is also hard for them to decipher the user’s original data from the stolen task-oriented semantic features.

Fig.~\ref {relay} illustrates an overview of Deep-JSCRC, with an unmanned aerial vehicle (UAV) photographing a \emph{sport utility vehicle (SUV)} and transmitting the image to a remote monitor without providing any additional description. Due to its low power, the UAV can only send the image to the nearby edge server (i.e., transmitter) that connects to another edge server (i.e., semantic relay) via the wireless channel. With the semantic complement of relays, additional semantic information, such as chassis height, is added to the updated semantic information. As a result, the monitor can correctly identify the \emph{SUV} using image semantics, whereas it is easily misidentified as a \emph{multipurpose vehicle (MPV)} using only its own background knowledge. 

\vspace{-0.2cm}
\subsection{{Collaborative Device-Server Optimization and Inference}}
Effective collaborative communications on large-scale devices are limited by diverse application scenarios, insufficient computing capability of terminal devices, data privacy leakage, and excessive communication latency. To address these issues, {collaborative device-server optimization and inference are proposed, including the following three aspects.}

\subsubsection{Distributed Adaptive Workload Optimization}
In practice, the computing resources required for inference are distinct and may change dramatically over time, making it difficult to guarantee the inference speed for heavily loaded devices. To address this problem, the distributed adaptive workload optimization strategy is proposed to adaptively transfer part of the load from high-loaded devices to low-loaded devices for devices in a compute cluster covered by the same edge server \cite{distream}. Specifically, the workload adjustment among devices can be calculated as $u_{i} = w_{i} + \sum\nolimits_{j}^{}{x_{ij}} + \sum\nolimits_{j}^{}{x_{ji}} + \hat{w_{i}}$, where $w_{i} $ denotes the workload of device $i$, $\hat{w_{i}} $ represents the upcoming workload of the device predicted by the Long Short-Term Memory (LSTM) based workload predictor, $\sum_{j}^{}{x_{ij}} $ and $\sum_{j}^{}{x_{ji}} $ indicate workload transferred from the each device respectively. In addition, the contribution of the semantic features transmitted by distributed devices to the task can be quantified by monitoring the effect of features on the gradient of the cross-entropy loss~\cite{gradient}.
Thus, devices with larger contributions can be allocated more computational and communication resources, resulting in efficient utilization of resources and lower latency. Hence, the adaptive optimization strategy can prevent workload accumulation on terminal devices, significantly reducing workload queuing time.

\subsubsection{Parameter Sharing Mechanism Based on Federated Learning}
To fully utilize the computational power of edge servers and distributed devices, a FL-based parameter sharing mechanism can be employed to improve  the training efficiency of the proposed DTCN model. Based on FL, the model parameters are learned by multiple edge servers or devices and the global server cooperatively. 
Specifically, the dual method and quadratic approximation can be used to obtain independent and learnable tasks when dealing with non-independent and identically distributed (non-IID) data from multiple local edge servers and devices. Then, each edge server or device can train its local model using the regionally collected data and transmit the updated local model parameters to the global server. A multi-device detection module computes the weighted average parameters of all local model parameters as the global model parameters and broadcasts them to all edge servers or devices. The communication rounds continue until the predefined convergence threshold is reached. In contrast to traditional centralized learning, in which all data is transmitted to the global server, the FL parameter sharing mechanism not only reduces communication load but also protects user privacy.
\subsubsection{Distributed Joint Device-Server Inference with Semantic Relay}
To reduce communication latency and workload, we develop a distributed joint device-server inference strategy with semantic relays that is divided into two phases, namely the local inference phase and the global inference phase. During the local inference phase, terminal devices transmit semantic features extracted from multimodal data using their own or adjacent devices' computing resources and offload them to the edge server. During the global inference phase, edge servers employ more complex DNNs to perform further inference. Specifically, the correlation between multimodal information is used to supplement and strengthen fused semantic features in order to suppress channel and semantic noise, facilitating the receiver to perform further inference easily. In contrast to the existing joint device-server inference in \cite{inference}, with the aid of semantic relay, terminal devices only need to deploy simple DNNs and transmit low-dimensional vectors, reducing computational resources and communication latency.


\section{Experimental Results and Discussion}
\label{fusionin}
In this section, we employ a multimodal classification task to demonstrate the effectiveness and robustness of the proposed DTCN against channel noise and modality absence.
\vspace{-0.3cm}
\subsection{Multimodal Classification Task with Semantic Relay} 
Multimodal classification is the common task of classifying target objects using different modality data, with image and text being the two most commonly used modalities. 
As a result, for simplicity, we implement the Deep-JSCRC scheme of DTCN for multimodal classification using image and text modality data. As shown in Fig.~\ref{relay}, the transmitter consists of a semantic encoder and a JSC encoder. The semantic encoder utilizes Inception-V3~\cite{inceptionv3}, whereas the JSC encoder employs multilayer DNNs. At the semantic relay, a JSC decoder employs multilayer DNNs, whereas the text semantic encoder employs a BERT cascaded LSTM to extract text modality features \cite{bert}. The additional semantic features provided by the text semantic encoder are fused with the semantic features of the transmitted images before being forwarded to the receiver via a JSC encoder. The fused multimodal features are recovered at the receiver using a JSC decoder and fusion semantic decoder, which are used to complete the multimodal classification task. We compare three baselines, namely single text modality (JSCC-T) and single image modality (JSCC-I and distributed information bottleneck variational feature encoding (DIB-VFE-I)\cite{DIB}), where Inception-V3 is added in front of DIB-VFE to simplify the preprocessing of the image of the following UPMC Food-101 dataset. Meanwhile, the performance upper bounds for multimodal and singlemodal are given, which are the classification accuracy of DTCN and JSCC-I in ideal noiseless channels, respectively.
\vspace{-0.3cm}
\subsection{Experimental Dataset and Settings}
In the experiment, we employ the UPMC Food-101 dataset\cite{dataset}, which includes $67,988$ training samples and $22,716$ test samples (covering food images and associated text recipe descriptions) divided into $101$ categories. Since images in different categories are similar or easily confused, the semantic features of the target image must be accurately recovered by the receiver to correctly identify the food category.

The Deep-JSCRC model is trained using both centralized learning and FL techniques to improve performance and training efficiency. Fig.~\ref{relay} shows the three-phase training of the Deep-JSCRC model utilizing centralized learning, namely, 1) training of semantic coding and decoding as well as semantic information fusion at the relay with cross-entropy loss; 2) training of JSC coding and decoding with $L_1$ loss; and 3) joint training of all modules with cross-entropy loss between the ground truth of input images and final categorized predictions. Meanwhile, the simulated channel is assumed to be an additive white Gaussian noise (AWGN) channel, and the signal-to-noise ratio (SNR) regimes are the same in the training and testing phases. 
Moreover, during the training process of FL, we consider that there are $10$ edge servers and a semantic relay as the global server. The UPMC Food-101 dataset is equally distributed to the edge servers. Besides, the number of training rounds is the same for both centralized learning and FL.

\begin{figure}[t]\vspace{-0.3cm}
	\centering
	{\label{Figgg}
		\includegraphics[width=0.83\columnwidth]{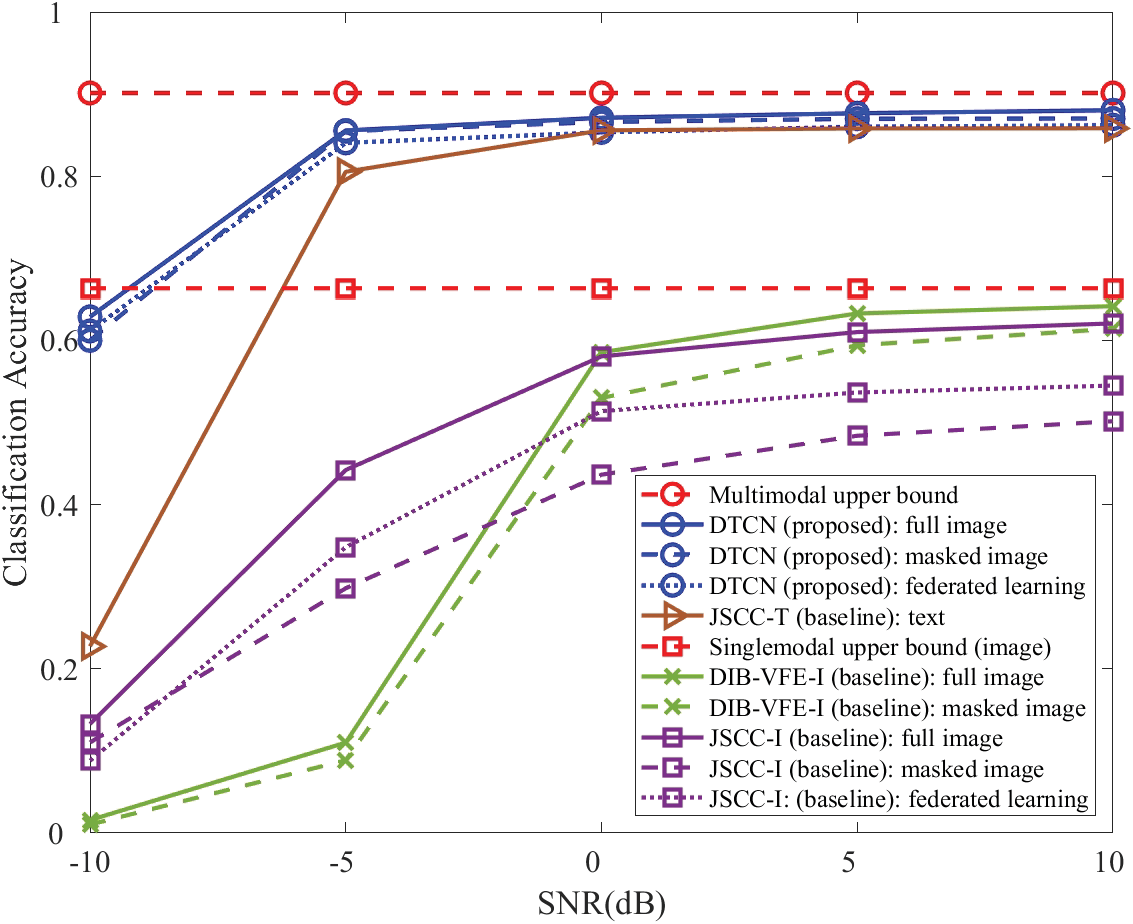}}\vspace{-0.2cm}
	\caption{Classification accuracy of the proposed DTCN , JSCC-T, JSCC-I, and DIB-VFE-I \cite{DIB} over AWGN channels.} 
	\label{exp_reslut}
\end{figure}

\vspace{-0.4cm}
\subsection{Comparison with Existing Methods}
Fig.~\ref {exp_reslut} shows that the proposed DTCN outperforms JSCC-T with single text modality, JSCC-I and DIB-VFE-I \cite{DIB} with single image modality, while approaching the multimodal upper bound when $\text{SNR} \ge 0$ dB. In the low SNR regime (i.e., $\text{SNR} = -10$ dB), the classification accuracies  of JSCC-T, JSCC-I, and DIB-VFE-I are as low as $22.7\%$, $13.27\%$, and $1.57\%$, respectively, while DTCN can still achieve $62.86\%$.
Remarkably, utilizing the complementary semantic information provided by semantic relays, the classification accuracy of the proposed DTCN is $20\%$ higher than the single-modal upper bound with images for $\text{SNR} \ge -5$ dB. As a result, the proposed DTCN can effectively suppress semantic noise and significantly improve classification accuracy, particularly in low SNR regimes.

Meanwhile, we consider an harsh communication environment to verify the robustness of the proposed DTCN, which is simulated by masking $50\%$ of the transmitted images. Fig.~\ref {exp_reslut} shows that there is only a slight performance degradation of the proposed DTCN with masked images. The reason is that the missing image information is compensated by complementary semantic features at the relay nodes. For JSCC-I and DIB-VFE-I with masked images, the classification accuracies decrease from $62.06\%$ and $64.17\%$  of full images to $50.15\%$ and $61.41\%$ of masked images at $\text{SNR}=10$~dB, respectively. This comparison verifies that DTCN can perform the classification task reliably even when some of the image modalities are absent due to poor channel conditions.

In addition, Fig.~\ref{exp_reslut} also shows that the proposed DTCN's performance loss in FL is within $2\%$ when compared to centralized learning. In comparison, the baseline JSCC-I suffers a performance loss of around $16.7\%$ in FL instead of centralized learning. This result demonstrates that the proposed DTCN can achieve high training efficiency with storng robustness in FL.

\begin{figure}[t]\vspace{-0.3cm}
	\centering
	\includegraphics[width=0.8\columnwidth]{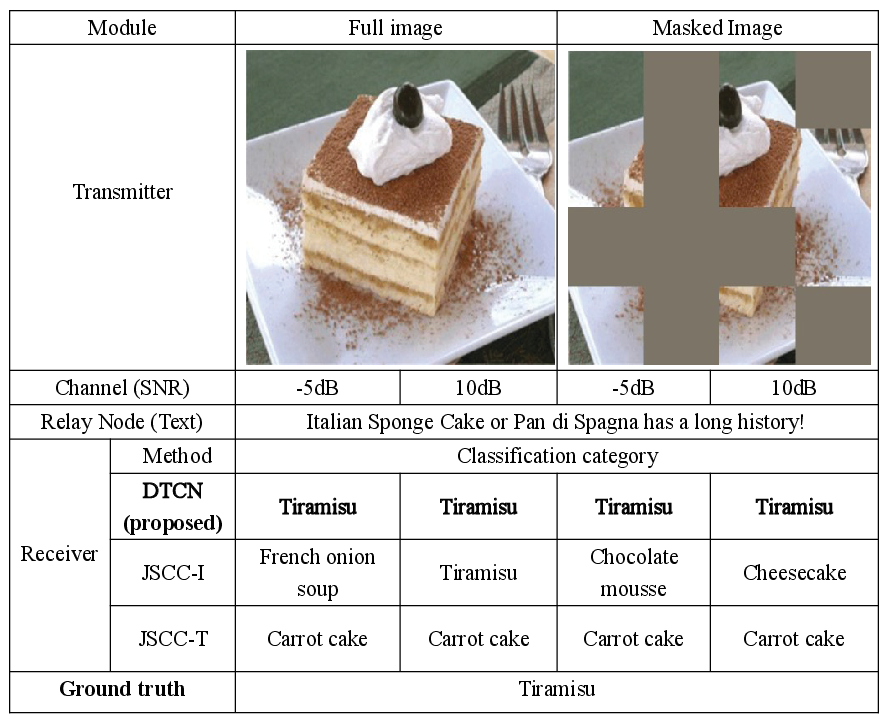}
	\caption{Visual results of the proposed DTCN, JSCC-I with full and masked images, and JSCC-T with text available at the relay.}
	\label{vir}
\end{figure}

Furthermore, Fig.~\ref{vir} illustrates the visual  comparison of DTCN, JSCC-I, and JSCC-T. Note that JSCC-I can only accurately identify the food category with full images and high SNR (i.e., $\text{SNR}=10$ dB). The JSCC-T also fails to correctly identify the food category based on the text available at the relay. In contrast, the DCTN can accurately identify the food category by fusing semantic information from text and image modalities, even in masked images with few valid pixels.

\vspace{-0.3cm}
\section{Conclusion}
\label{conclus}
This article proposes a novel DTCN framework based on multimodal semantic relay and edge intelligence. To realize DTCN, we provide the related fundamental techniques, such as semantic alignment and completion, Deep-JSCRC, and collaborative device-server optimization and inference. The keys of these techniques are to explore and fuse the underlying knowledge and correlation from the multimodal data based on the task requirement. Furthermore, the common multimodal classification task is used as an example to validate the effectiveness and robustness of the proposed DTCN.


\end{document}